\documentstyle[amssymb,12pt,epsf,palatino]{nature-pvd}

\def\simlt{\mathrel{\hbox{\rlap{\hbox{\lower4pt\hbox{$\sim$}}}\hbox{$<$}}}}
\def\simgt{\mathrel{\hbox{\rlap{\hbox{\lower4pt\hbox{$\sim$}}}\hbox{$>$}}}}

\def\ale{\mathrel{\hbox{\rlap{\hbox{\lower4pt\hbox{$\sim$}}}\hbox{$<$}}}}
\def\age{\mathrel{\hbox{\rlap{\hbox{\lower4pt\hbox{$\sim$}}}\hbox{$>$}}}}

\def\kms{km\,s$^{-1}$}
\def\msun{M$_{\odot}$}

\def\spose#1{\hbox to 0pt{#1\hss}}
\newcommand\lsim{\mathrel{\spose{\lower 3pt\hbox{$\mathchar''218$}}
     \raise 2.0pt\hbox{$\mathchar''13C$}}}
\newcommand\gsim{\mathrel{\spose{\lower 3pt\hbox{$\mathchar''218$}}
     \raise 2.0pt\hbox{$\mathchar''13E$}}}

\newcommand{\dg}{\mbox{$^\circ$}}
\newcommand{\am}{\mbox{$^{\prime}$}}
\newcommand{\as}{\mbox{$^{\prime\prime}$}}

\newcommand{\twco}{\mbox{$^{12}\textrm{CO}$}}
\newcommand{\thco}{\mbox{$^{13}\textrm{CO}$}}
\newcommand{\ceio}{\mbox{$\textrm{C}^{18}\textrm{O}$}}

\begin{document}
\gdef\ha{H$\alpha$}
\gdef\ew{${\rm EW}({\rm H}\alpha)$}
\gdef\msun{M$_{\odot}$}
\gdef\kms{km\,s$^{-1}$}


\title{\Large \bf {Episodic molecular outflow in the very young protostellar cluster Serpens South}}

\author{{Adele L.~Plunkett}\affiliation[1]
  {{Astronomy Department, Yale University,  New Haven, CT 06511, USA}
},
   {{H\'{e}ctor G.~Arce}}$^1$,   
   {Diego Mardones}\affiliation[2]
   {{Departamento de Astronom\'{i}a, Universidad de Chile, Casilla 36-D, Santiago, Chile}},  
   {{Pieter van Dokkum}}$^1$,   
 {Michael M.~Dunham}\affiliation[3]
 {{Harvard-Smithsonian Center for Astrophysics, 60 Garden Street, MS 78, Cambridge, MA 02138, USA}},  
 {Manuel Fern\'{a}ndez-L\'{o}pez}\affiliation[4]
{{Instituto Argentino de Radioastronom\'{i}a, CCT-La Plata (CONICET), C.C.5, 1894, Villa Elisa, Argentina}},  
 {Jos\'{e} Gallardo}\affiliation[5] 
{{Joint ALMA Observatory, Av. Alonso de C\'{o}rdova 3107, Vitacura, Santiago, Chile}},  
 {Stuartt A.~Corder}$^5$ 
   {}\vspace{0.4cm}
}
\headertitle{Episodic outflow in Serpens South}
\mainauthor{}

\summary{ACCEPTED FOR PUBLICATION IN \emph{NATURE}.  The loss of mass from protostars, in the form of a jet or outflow, is a necessary counterpart to protostellar mass accretion\cite{Nor80,Shu87}.  Outflow ejection events probably vary in their velocity and/or in the rate of mass loss. Such `episodic' ejection events\cite{Fra14} have been observed during the Class 0 protostellar phase (the early accretion stage)\cite{Gue99,Lee00,Lee07a,Lee07b,San09,Hir10,Loi13}, and continue during the subsequent class I phase that marks the first one million years of star formation\cite{Cab00,Goo04,Ioa12,Arc13}.  Previously observed episodic-ejection sources were relatively isolated; however, the most common sites of star formation are clusters\cite{Lad03}. Outflows link protostars with their environment and provide a viable source of turbulence that is necessary for regulating star formation in clusters\cite{Fra14}, but it is not known how an accretion-driven jet or outflow in a clustered environment manifests itself in its earliest stage.  This early stage is important in establishing the initial conditions for momentum and energy transfer to the environment as the protostar and cluster evolve.  Here we report that an outflow from a very young class 0 protostar, at the hub of the very active and filamentary Serpens South protostellar cluster\cite{Gut08,Tan13,Nak14sers}, shows unambiguous episodic events.  The $^{12}$C$^{16}$O ($J=2-1$) emission from the protostar reveals 22 distinct features of outflow ejecta, the most recent having the highest velocity.  The outflow forms bipolar lobes --- one of the first detectable signs of star formation --- which originate from the peak of 1-mm continuum emission. Emission from the surrounding C$^{18}$O envelope shows kinematics consistent with rotation and an infall of material onto the protostar.  The data suggest that episodic accretion-driven outflow begins in the earliest phase of protostellar evolution, and that the outflow remains intact in a very clustered environment, probably providing efficient momentum transfer for driving turbulence.
}
\maketitle
%

\onecolumn 

\vspace{-0.0cm}
\noindent

We used the Atacama Large Millimeter/sub-millimeter Array (ALMA) in Chile to observe the $J=2-1$\ emission line of carbon monoxide isotopologues ($^{12}$CO, $^{13}$CO and C$^{18}$O) near the class 0 source CARMA-7 (hereafter C7), in the young protostellar cluster Serpens South.  C7 is the strongest of several millimetre-wavelength continuum sources that are densely packed within Serpens South, located at a distance of 415 parsec (pc) from Earth\cite{Dzi10}. Its relative proximity to Earth allows for observations with high spatial resolution; our observations resolve features with physical sizes of greater than about 370 astronomical units (AU).  

\begin{figure*}[!ht]
\epsfxsize=15cm
\epsffile{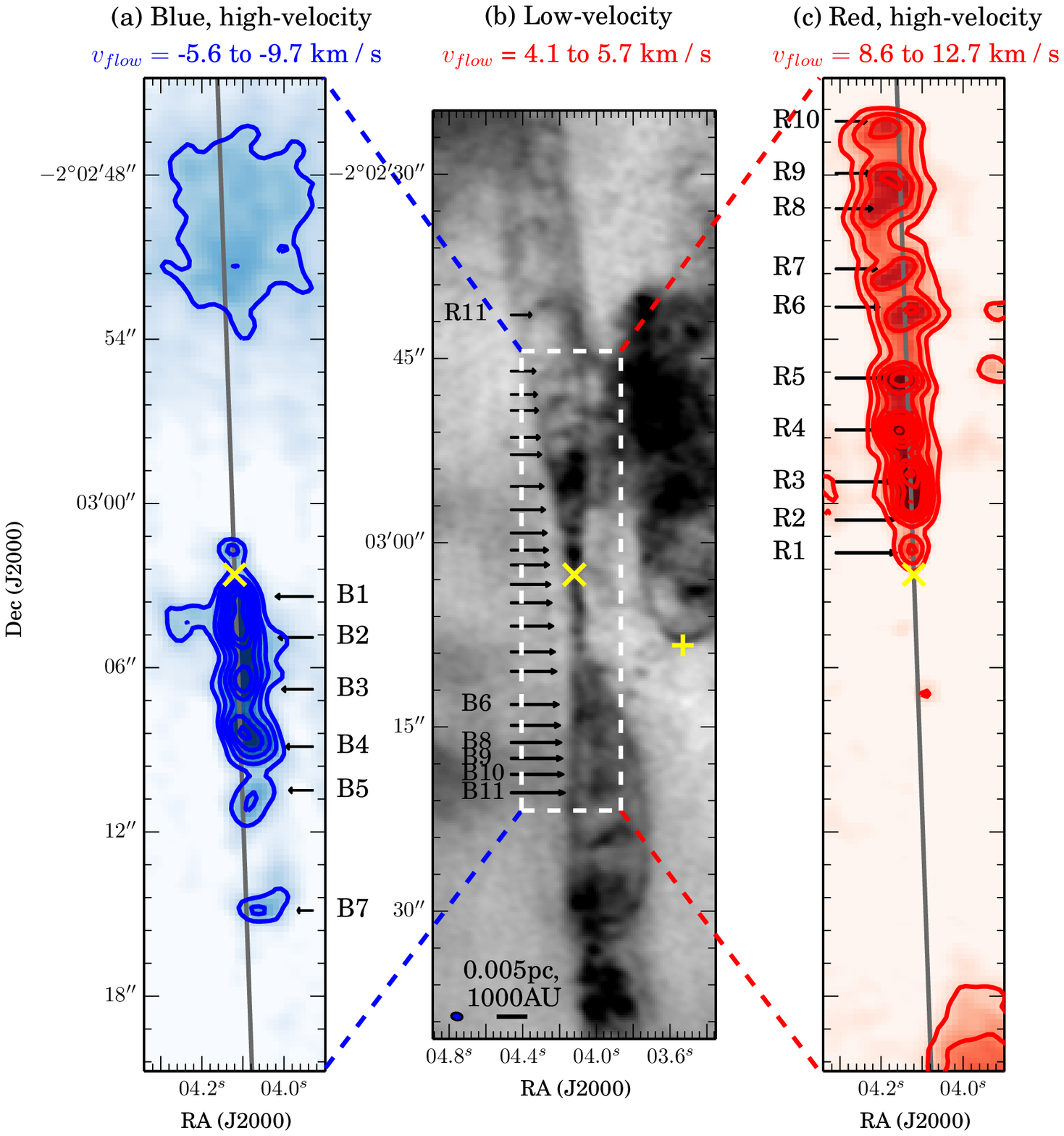}\\
{\small \bf {\textsf{Figure 1}} $|$ } 
{\small \bf {\textsf{\twco\ molecular outflow emission centered at the class 0 protostar CARMA-7 (C7).}}}
{\small
C7 is marked by the yellow cross at RA $=$ 18 h 30 min 04.1 s, declination dec. $=-02\dg\ 03\am\ 02.6\as$. The numbers on the $x$\ axes are truncated to show seconds only, omitting hours and minutes for brevity.  The $y$\ axes are likewise simplified. \textbf{a, c,} High-velocity blueshifted and redshifted channels, respectively.  \textbf{b,} Low-velocity channels, to show the cavity surrounding collimated ejecta.  Contours in \textbf{a} and \textbf{c} begin with $8\sigma$\ and increment by $4\sigma$\ and $8\sigma$, respectively.  Labels B1--B11 and R1--R11 indicate 22 ejecta features.  The gray line marks the 4\dg\ position angle of the C7 outflow lobes.  The yellow `plus' symbol marks a neighbouring protostar, CARMA-6\cite{Plu15}, which provides contaminating emission, especially for the blueshifted southern outflow lobe.  
}
\label{fig:panel_12co}
\end{figure*}

\begin{figure*}[!ht]
\epsfxsize=15cm
\epsffile{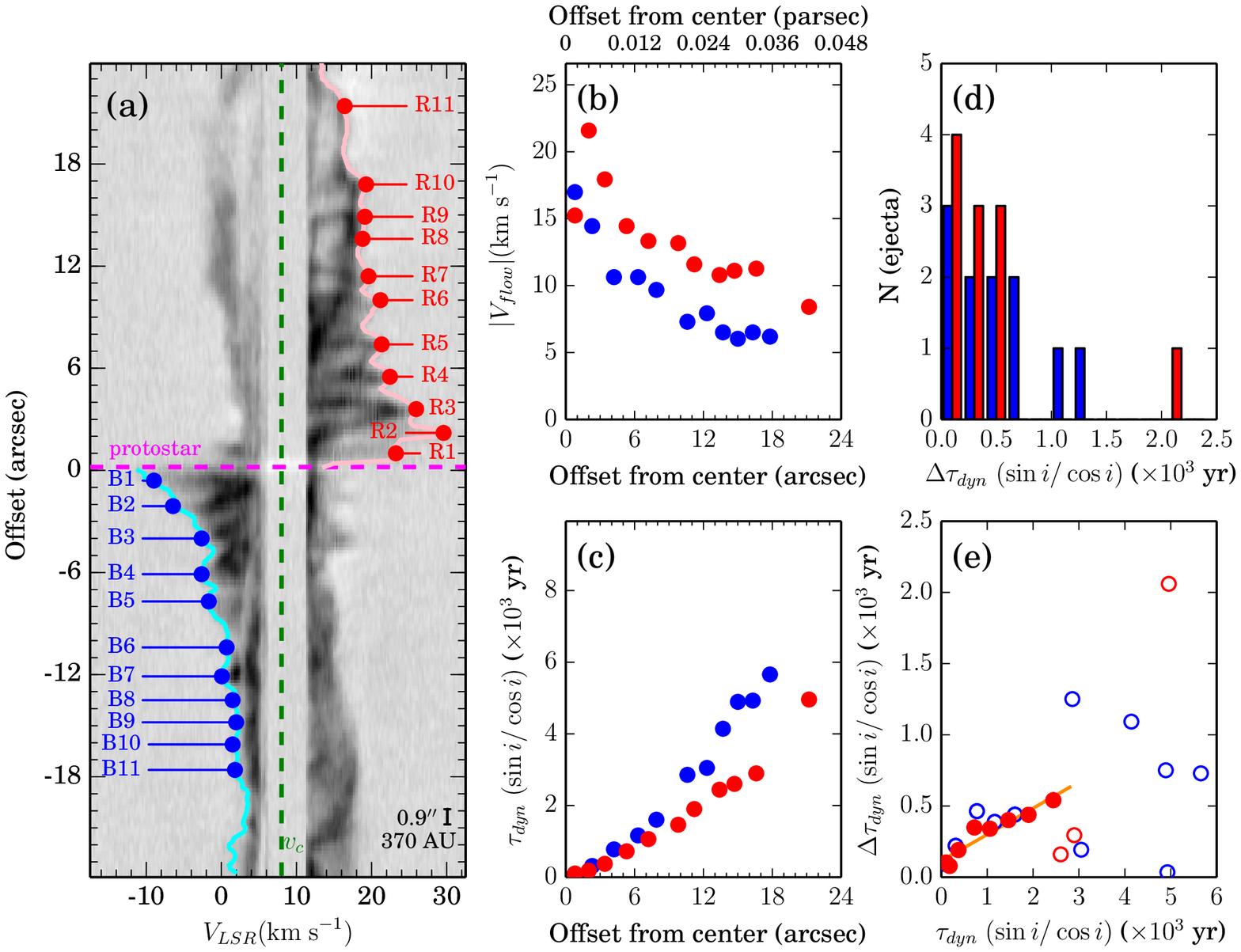}\\
{\small \bf {\textsf{Figure 2}} $|$ }
{\small \bf {\textsf{Outflow ejecta from C7.}}}
{\small
\textbf{a,} Position--velocity diagram along the outflow-axis (see Fig.~1\ref{fig:panel_12co}).  Points correspond to velocity maxima where we identify outflow knots.  Northern emission features are mostly redshifted and are shown in red; southern emission features are mostly blueshifted and are shown in blue.  The scale bar shows 370 AU, or 0.9\as.  The dashed pink line marks the location of the protostar; the dashed green line snows the cloud central velocity, $V_C$, in the same units as those of $V_{LSR}$.  \textbf{b,} Knot velocity ($V_{flow}$) versus distance relative to C7 (in arcsecs or parsecs). Blue (red) points mark southern (northern) features, as in \textbf{a}. \textbf{c,} Dynamical timescales ($\tau_{dyn}$) for each knot, with no correction for inclination angle. \textbf{d,} Histogram showing the number ($N$) of ejecta that have been emitted at the given times since the previous ejection ($\Delta \tau_{dyn}$), with 200-year bins. \textbf{e} $\Delta \tau_{dyn}$ as a function of $\tau_{dyn}$ for northern (red) and southern (blue) knots.  Recent northern ejecta (solid points) are fit with a linear trend (orange line). 
}
\label{fig:pv}
\end{figure*}

The \twco\ emission extends north--south of C7, spanning about $80\as$\ (or 0.16 pc) along an axis with a position angle of roughly $4\dg$\ (Fig.~1\ref{fig:panel_12co}).  The emission is clumpy, and the strongest emission features to the north (south) are mostly redshifted (blueshifted), relative to the systemic cloud velocity ($V_{c}$) of 8\ \kms.\cite{Kir13,Plu15}  The emission features near the origin are only around $1-2\as$\ (about $400-800$\ AU) wide, and the width increases to about $8\as$\ (roughly $3000$\ AU) at the widest point.  The opening angle of the emission decreases with velocity, with a maximum of about $23\dg$\ (at $10\as$, or 0.02 pc, from the source) at velocities of a few kilometres per second, and a minimum of about $10\dg$ at the same distance and higher velocities.  Figure 2\ref{fig:pv}a shows the position--velocity diagram, with a saw-like pattern along the extent of the \twco\ emission; and emission features corresponding to the highest velocities ($|V_{LSR}-V_{c}|=\sim 20$\ \kms, where $V_{LSR}$\ is the local standard of rest velocity) are found closest to C7 (Fig.~2\ref{fig:pv}a, b).  The \twco\ emission is optically thick --- especially near the cloud velocity according to our data --- and therefore it traces outflow features with velocities greater than a few kilometres per second with respect to $V_{c}$.   

\begin{figure*}[!ht]
\epsfxsize=15cm
\epsffile{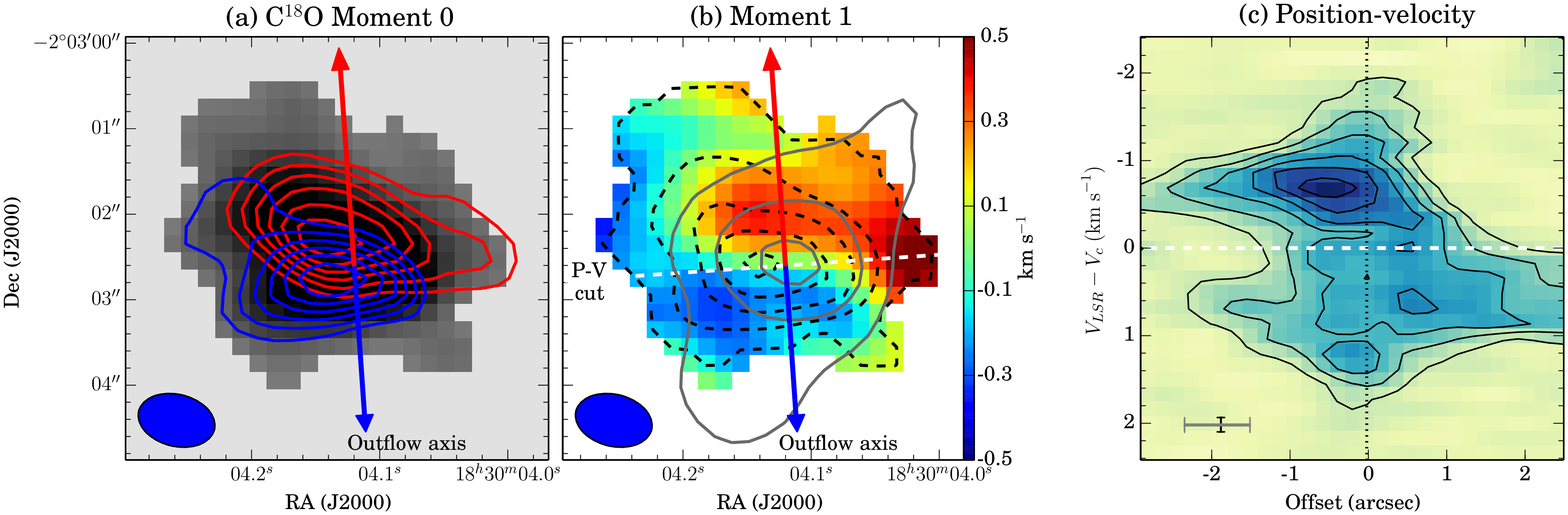}\\
{\small \bf {\textsf{Figure 3}} $|$ }
{\small \bf {\textsf{Protostellar envelope. }}}
{\small
\textbf{a,} Integrated \ceio\ intensity (moment 0, grayscale) $>13\sigma$ with $|v_{LSR}-v_c|<3$\ \kms.  Blue (red) contours represent blueshifted (redshifted) channels, with $0.4<|v_{LSR}-v_c|<3$\ \kms, beginning at 30\% of peak integrated intensity (165 mJy beam$^{-1}$\ \kms\ and 155 mJy beam$^{-1}$\ \kms, respectively) with increments of 10\% of peak.  The blue oval represents the beam size. \textbf{b,} Intensity-weighted mean velocity (moment 1, colour scale).  Black dashed contours show integrated intensity (greyscale in \textbf{a}) with $8\sigma$\ steps.  Grey contours show 15\%, 30\%, and 80\% of peak continuum emission (93.9 mJy beam$^{-1}$).  \textbf{c,} Position--velocity perpendicular to the outflow axis. Contours show levels of $4\sigma$\ of the position--velocity intensity. Spatial and velocity resolution elements are shown with grey and black (solid) bars, respectively.
}
\label{fig:pv_c18o}
\end{figure*}

\begin{figure*}[!ht]
\epsfxsize=15cm
\epsffile{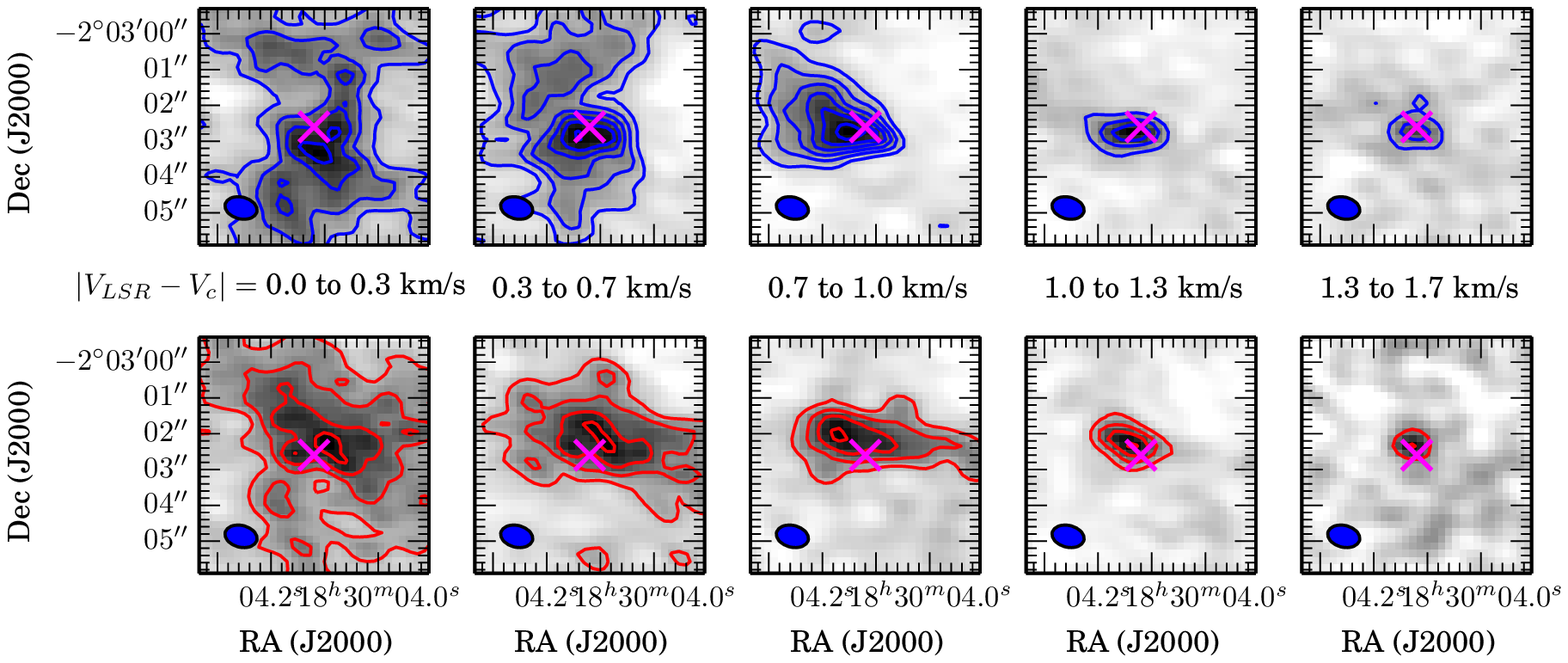}\\
{\small \bf {\textsf{Extended Data Figure 1}} $|$ }
{\small \bf {\textsf{\ceio\ emission from the protostellar source C7.}}}
{\small
Top row, blueshifted emission; bottom row, redshifted emission; velocity increases from left to right. Contours begin at $4\sigma$\ and increment by $4\sigma$\ for each respective integrated intensity map. Specific velocity ranges ($V_{LSR}-V_c$, or velocity relative to cloud velocity) are given
for each column, with each panel showing integrated emission from two channels. The location of peak continuum emission is marked with a magenta cross.}
\label{fig:channel_c18o}
\end{figure*}

\begin{figure*}[!ht]
\epsfxsize=15cm
\epsffile{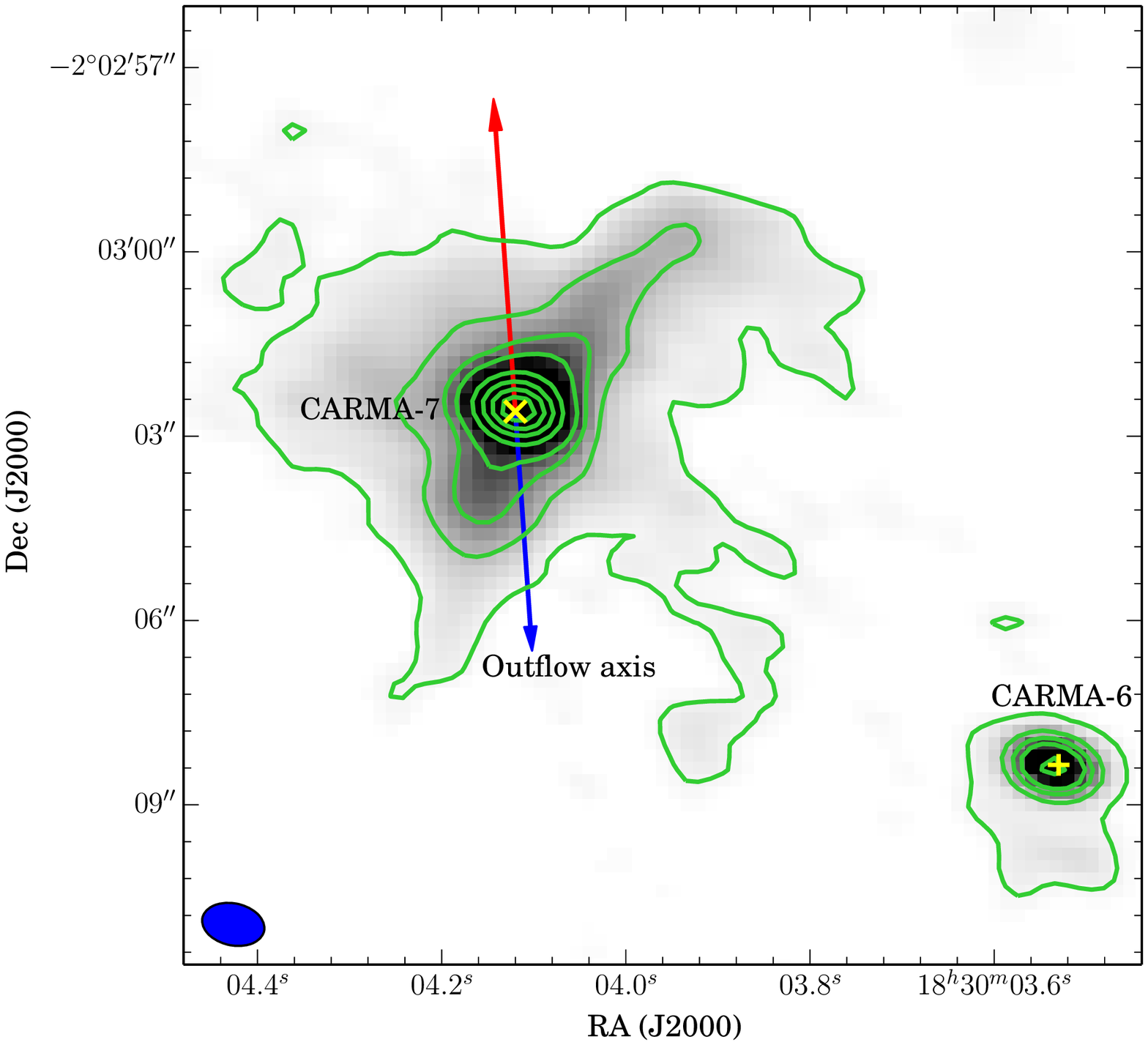}\\
{\small \bf {\textsf{Extended Data Figure 2}} $|$ }
{\small \bf {\textsf{1-mm continuum emission near the sources CARMA-7 (RA $=$ 18 h 30 min 04.1 s, dec. $=-02\dg\ 03\am\ 02.6\as$) and CARMA-6 (RA $=$ 18 h 30 min 03.5 s, dec. $=-02\dg\ 03\am\ 08.4\as$). }}}
{\small
Contours show 10$\sigma$, 30$\sigma$, 50$\sigma$ and 70$\sigma$, followed by increments of 50$\sigma$.  Near these strong sources, we find the r.m.s. noise to be 0.3 mJy beam$^{-1}$.
}
\label{fig:contin}
\end{figure*}

The \ceio\ emission is optically thin and therefore probes deeper than does the \twco\ emission, to trace denser material that is closer to the protostar (see Fig.~3\ref{fig:pv_c18o} and the channel maps in Extended Data Fig.~1\ref{fig:channel_c18o}).  Together, these molecular lines and continuum (Extended Data Fig.~2\ref{fig:contin}) trace two related components of the protostellar system: the outflow and the envelope.  Material accretes onto a protostar from an infalling envelope via a disk, with the envelope providing the main mass reservoir for the star.  While the protostar is still obscured by the surrounding envelope, a bipolar outflow represents one of the first observational signs of star formation, and it carries away mass and angular momentum from the system.  

Our observations show two molecular outflow lobes emanating from the C7 envelope, and we conclusively identify an outflow-driving source in this region.  When this region was studied with lower resolution CO observations\cite{Nak11sers,Plu15}, prevalent outflow emission from several young sources appeared to coincide.  However, the \twco\ emission traces cool (less than about $100$\ K) swept-up outflow material and provides a record for the timing history of mass-loss events for a given source.  The C7 outflow comprises cavity walls that surround 22 knots (observed clumps of emission from a single ejection event), 11 to the north and 11 to the south, within $24\as$ from the source.  Beyond this distance, we see outflow morphology that can be attributed to C7, but there is contaminating cloud emission to the north and an interfering outflow to the south (driven by a protostar southwest of C7; Extended Data Fig.~2\ref{fig:contin}).  The outflow's high collimation, and the presence of redshifted and blueshifted emission coinciding along the line of sight near the protostar north and south, are consistent with the main outflow axis oriented nearly in the plane of the sky.  Low-velocity redshifted and blueshifted emissions to the south and north are contributed by cavities surrounding the high-velocity jet-like emission, and the jet may precess slightly, given the slight wiggle in the knots shown in Fig.~1\ref{fig:panel_12co}a, c.

Clumpy \twco\ emission suggests an episodic ejection mechanism, rather than a smooth outflow. Decreasing knot velocities with distance from C7 are consistent with the existence of jet-entrained material that is slowed down by drag because of interaction with the surrounding medium, and/or with the existence of intrinsically variable ejections\cite{Rag90,Sut97,Smi97}; both probably contribute to the position--velocity trend seen here.  The  `superjet' HH34 (driven by the class I source HH34 IRS) also shows a velocity decrease\cite{Cab00}, which is proposed to be caused by the drag-induced slowing of jet-entrained material.  However, the shapes of the position--velocity curves for HH34 and C7 differ, a difference that may be explained by the relative ages and precession of the sources.  The initial C7 ejecta probably cleared some of the dense ambient material, reducing the drag forces for later ejecta following closely behind and in line with previous ejecta.  HH34 is more evolved and precessing to a greater extent, so ejecta seem to be more directly exposed to ambient material, which has not yet been disturbed by previous ejections.

We also find that the velocities of southern (blueshifted) knots from C7 are consistently lower than the velocities of northern (redshifted) knots at comparable distances.  This may be evidence for an inhomogeneous ambient cloud medium, such that the southern knots are being slowed down by a denser environment.  Alternatively, the jets may be intrinsically variable upon ejection from opposite sides of the disk.  It is also possible that the outflow lobes have different inclination angles with respect to the plane of the sky, so that the line-of-sight velocities to the north and south differ.  C7 may precess slightly, since blueshifted emission near C7 shifts to being predominantly red farther from the source. 

In Fig.~2\ref{fig:pv}c we show dynamic timescales for each of the identified ejecta, ranging from 100 years to 6,000 years (for knots within $24\as$, or 10,000 AU, from the source).  The dynamic timescale for each ejection is given by $\tau_{dyn}=D/V_{flow}(\cos i/\sin i)$, where $D$\ is the distance between an outflow knot and the driving source, $V_{flow}$ is the velocity (along the line of sight) of the knot, and $i$\ is the inclination of the outflow with respect to the line of sight.  Uncertainties arise because we do not know the inclination angle, and because we assume that the knots travel with constant velocity from the time of their launch.  If a jet is launched from the disk\cite{Fra14}, then the longest timescale of an (unimpeded) outflow ejection should be a lower limit for the formation time of the disk.  The longest timescale of a northern ejection is about 5,000 years; correcting for an inclination angle nearly in the plane of the sky, this could be smaller by a factor of about 10 (for $i=\sim85\dg$) or more, which is consistent with the youthfulness of the source.  Given that southern knots appear to have lower velocities than northern knots, the timescales of the southern (blueshifted) knots are longer on average than the northern (redshifted) knots.

We quantify the episodic nature of the ejections, and corresponding accretion and/or disk instabilities\cite{Fra14,Aud14}, on the basis of the difference in timescales, $\Delta \tau_{dyn}$, for successive ejection events (Fig.~2\ref{fig:pv}d).  Because of the contamination from the surrounding outflow emission to the south, we base the following calculations on the northern lobe only (within 24\as\ of C7).  In Fig.~2\ref{fig:pv}e, we see that seven knots to the north show linearly increasing $\Delta \tau_{dyn}$\ as a function of $\tau_{dyn}$, with $\Delta \tau_{dyn}$ ranging from 80 years to 540 years, and mean $\Delta \tau_{dyn}$ of $310\pm150$\ years.  These seven knots are the most recently ejected to the north, with $\tau_{dyn}$\ of less than 2,400 years (uncorrected for inclination angle). Several modes of velocity variability have been suggested for protostellar jets\cite{Rag02,Ioa12}, with periods of a few tens, a few hundreds, and a few thousands of years; in the case of a class 0 source, and assuming that C7 has an inclination approximately in the plane of the sky, we are probably witnessing ejecta that are associated mostly with the shorter period modes.  

We estimate that, within some 3,000 years, the farthest (slowest) ejecta in the north and south will have been overcome by each of the following (faster) ejecta, if ejecta travel with constant velocities (an admittedly simple assumption). These interactions will produce bright shocks along the outflow.  `Snapshots' of shocks in outflows can be seen in the emission of molecular hydrogen (H$_2$)\cite{Tei12} emission, which has a higher excitation temperature than does \twco\ but cools quickly.  Two H$_2$\ bow-shaped shock structures, corresponding to faint, low-velocity \twco\ emission lines 38\as\ (0.08 pc) and 47\as\ (0.09 pc) north and south of C7, respectively, are seen in the Spitzer 4.5-$\mu$m map of the region\cite{Gut08}.  These structures may be evidence of the first occurrence of a longer-period mode (of a few hundred years or more), where faster ejecta recently overcame slower ejecta.  We propose that frequent ejection bursts during the class 0 phase entrain molecular outflow material, which therefore appears clumpy, creating observable shocks when the ejecta overtake previous ejecta.

Alternatively, if the position--velocity trend provides evidence for an interaction between ejecta and the environment, then the drag-induced momentum loss along the outflow signifies momentum transfer to the environment --- an important mechanism that is proposed to drive turbulent motions in a clustered region\cite{Fra14}.  We are carrying out further analysis of momentum injection along the span of the outflow at such an early stage, taking into account velocity-dependent opacity of the \twco\ line and varying excitation temperatures throughout the outflow.
 
The \ceio\ envelope seems to be oriented perpendicular to the outflow axis, with its major axis approximately east--west.  Elongated blueshifted and redshifted \ceio\ emissions east and west of C7, respectively, are evidence of a non-spherical, rotating envelope.  Blueshifted and redshifted peaks of high-velocity emission near C7 to the south and north, respectively, are consistent with infall motion onto a disk that is slightly inclined\cite{Oya14}. 

\begin{figure*}[!ht]
\epsfxsize=15cm
\epsffile{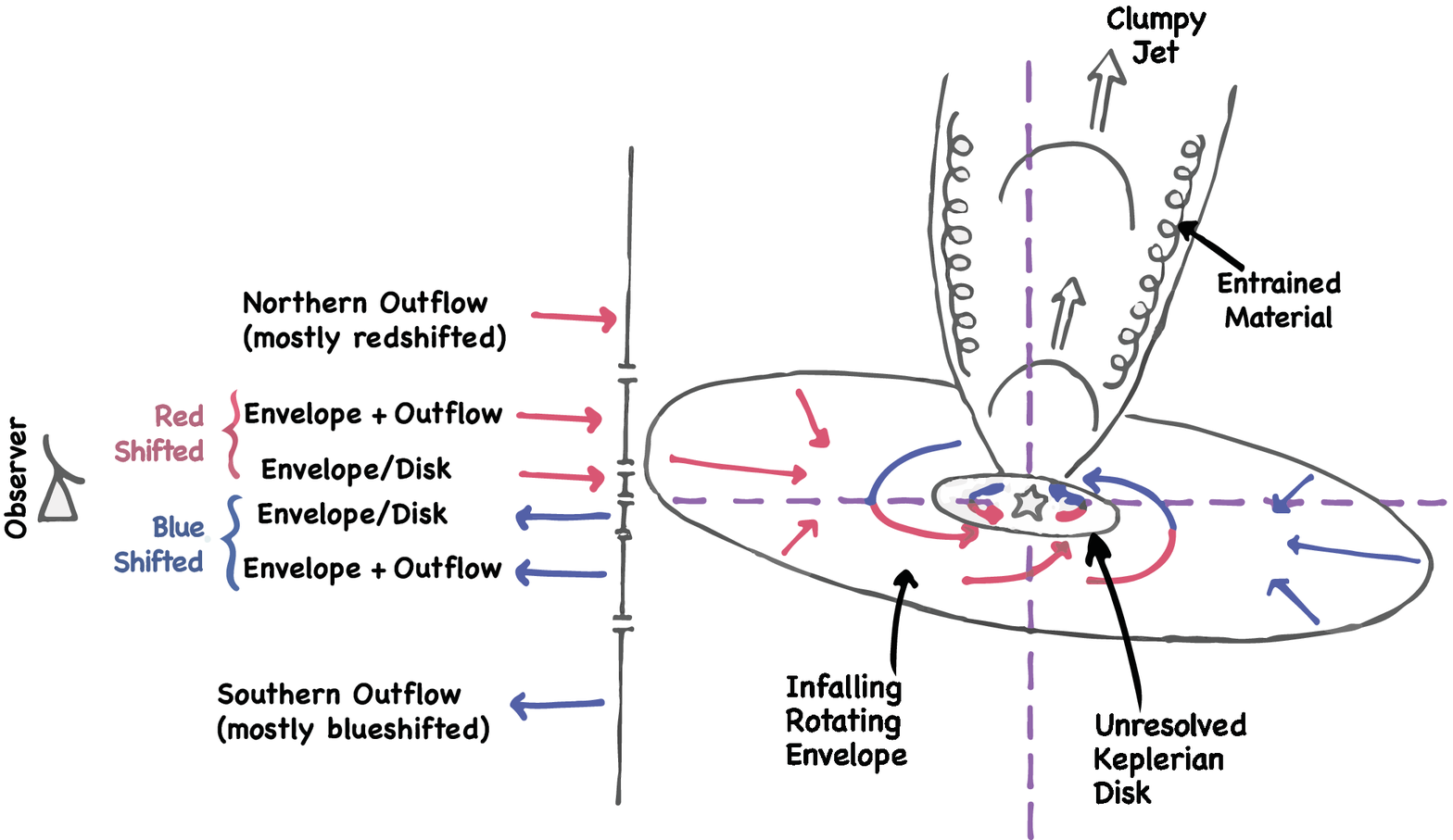}\\
{\small \bf {\textsf{Extended Data Figure 3}} $|$ }
{\small \bf {\textsf{Cartoon depiction of protostellar system showing the outflow (\twco\ emission), envelope (\ceio\ emission) and disk (unresolved).}}}
{\small Contributions to blueshifted and redshifted molecular line emission are indicated along the outflow and envelope, assuming the outflow is nearly in the plane of the sky with respect to the observer.  
}
\label{fig:cartoon}
\end{figure*}

Two features in the \ceio\ PV diagram are representative of some contribution from unresolved Keplerian rotation (on scales of less than $\sim400$\ AU): larger velocities at smaller distances, and position--velocity intensity peaks offset bluewards and redwards from the line $V_{LSR}=V_{c}$.  The position--velocity structure for C7 is consistent with a combination of rotation and infall on a slightly inclined disk, as shown in models\cite{Oya14} and sketched in Extended Data Fig.~3\ref{fig:cartoon}.  However, the \ceio\ position--velocity diagram (Fig.~3\ref{fig:pv_c18o}c) also shows some deviations from models of a rotating, infalling envelope: first, the blueshifted peak is stronger than the redshifted peak; second, redshifted emission with velocities $|V_{LSR}-V_{c}|\sim0.5-1$\ \kms\ coincides with strong blueshifted emission at an offset of about $-2\as$; and third, redshifted extended emission west of C7 probably contaminates the C7 envelope emission. The outflow and envelope that we observe here clearly pertain to the same protostar, and higher-resolution observations of the disk and envelope will reveal the jet-launching region and disk formation mechanisms in this young system.
\vspace{1cm}

\clearpage

\bibliographystyle{nature-pap}

\vspace{0.3cm}

\vspace{0.3cm}
\noindent
{\small \bf \textsf{Acknowledgements} }
{\small A.L.P. is supported by the National Science Foundation (NSF) Graduate Research Fellowship under Grant DGE-1122492; this research was made possible by the US Student Program of Fulbright Chile.  H.G.A.~receives funding from the NSF under grant AST-0845619.  D.M.~acknowledges support from CONICYT project PFB-06.  M.M.D.~acknowledges support from the Submillimeter Array through a postdoctoral fellowship.  ALMA is a partnership of the European Southern Observatory (ESO, representing its member states), NSF (USA) and National Institutes of Natural Sciences (Japan), together with the National Research Council (Canada) and National Security Council and Academia Sinica Institute of Astronomy and Astrophysics (Taiwan), in cooperation with the Republic of Chile. The Joint ALMA Observatory is operated by ESO, Associated Universities Inc.~(AUI)/National Radio Astronomy Observatory (NRAO) and National Astronomical Observatory of Japan.  The NRAO is a facility of the NSF, operated under cooperative agreement by AUI. This paper makes use of the following ALMA data: ADS/JAO.ALMA \#2012.1.00769.S.  }

\vspace{0.3cm}
\noindent
{\small \bf \textsf{Author Contributions}}
{\small A.L.P.~led the proposal, observations, analysis and interpretation, and wrote the manuscript. H.G.A.~contributed to the analysis and interpretation, and contributed to the manuscript.  A.L.P., H.G.A., D.M., M.M.D., J.G., and S.A.C.~planned the early stages of the project. D.M., M.M.D., M.F., and J.G.~contributed to the analysis and interpretation and commented on the manuscript. P.v.D.~contributed to the interpretation and to the manuscript.}

\vspace{0.3cm}
\noindent
{\small \bf \textsf{Author Information}}
{\small Reprints and permissions information is available at www.nature.com/reprints.  The authors declare no competing financial interests.  Readers are welcome to comment on the online version of the paper.  Correspondence and requests for materials should be addressed to A.L.P.~(aplunket@eso.org) or H.G.A.~(hector.arce@yale.edu).   }

\clearpage
\noindent
\noindent\textbf{METHODS}
\vspace{0.05cm}\\
\textbf{Observations and data analysis.} The analysis is based on ALMA Cycle 1 observations made with the 12-metre and 7-metre arrays during March 2014 and January to June 2014.  The observed mosaics span $2\am\times3\am$ and consist of 137 and 53 pointings separated by 15\as\ and 26\as\, made by the 12-metre and 7-metre arrays, respectively.  Here, we focus on the roughly $90\as\times20\as$\ region centered at RA $=$ 18 h 30 min 04.1 s, dec. $=-02\dg\ 03\am\ 02.6\as$.  

The ALMA correlator was configured in the frequency division mode (FDM) of Band 6 with four independent spectral windows: one window was assigned to the $J=2-1$\ energy level transition of each of the spectral lines \twco\ (230.538 GHz), \thco\ (220.399 GHz) and \ceio\ (219.560 GHz), and the fourth was dedicated to continuum at 231.450 GHz.  The bandwidth for each spectral-line window was 234.375 MHz, and the continuum window had a bandwidth of 468.750 MHz. To make a continuum-emission map, we included line-free channels in all spectral windows, resulting in a total continuum bandwidth of 996 MHz. The molecular line data for \twco\ and \ceio, as well as the continuum, are included in the present analysis.

We performed calibration of the raw visibility data with the Common Astronomy Software Application (CASA, version 4.3.0) using the standard reduction script for Cycle 1.  We assigned weights to the measurement sets using the task `statwt' and combined the calibrated 12-metre and 7-metre array uv-data using the task `concat'.  

We created image cubes for each molecular line, as well as the continuum image, by first applying a Fourier transform to the calibrated data, producing an intermediate (`dirty') image.  Using the intermediate image, we drew masks around the emission features, and these masks were used in an interactive `clean' process to deconvolve the telescope point-spread function.  We used Briggs weighting with a robust parameter of 0.5, and we imaged with a cell size of 0.3\as\ and a spectral (velocity) resolution of 0.16 \kms.  Finally, we subtracted continuum emission from the spectral line data using the task `imcontsub'.

The resulting beam sizes for the \twco\ and \ceio\ data cubes are $0.9\as\times0.6\as$ (with position angles of $79.7\dg$\ and $76.3\dg$\ for \twco\ and \ceio, respectively).  The root-mean-squared (r.m.s.) noise levels are 9 mJy beam$^{-1}$\ channel$^{-1}$\ and 8 mJy beam$^{-1}$\ channel$^{-1}$\, respectively, with channel widths of 0.16 \kms.  The r.m.s. noise level for the continuum is 0.2 mJy beam $^{-1}$\ near the edge of the region presented here, with an upper-limit r.m.s. noise level of 0.3 mJy beam$^{-1}$\ within 30\as of the strong continuum emission.

\noindent \textbf{\ceio\ channel maps.} The \ceio\ emission, shown in Extended Data Fig.~1\ref{fig:channel_c18o}, is concentrated where the northern redshifted and southern blueshifted \twco\ emissions meet.    The \ceio\ morphology changes from extended at the lowest velocities ($|V_{LSR}-V_{c}|=0-0.7$\ \kms) to compact and oriented approximately north--south, or coincident with the outflow axis, at higher velocities ($|V_{LSR}-V_{c}|=1.3-1.7$\ \kms).  At intermediate velocities ($|V_{LSR}-V_{c}|=0.6-1$\ \kms), elongated blueshifted and redshifted emissions are seen east and west of C7, respectively.  A shell in the \ceio\ emission in the south, and less significantly in the north, is seen bisected by the \twco\ axis in Fig.~3\ref{fig:pv_c18o}a, b.  This is similar to the situation with the protostar HH212\cite{Lee14}, and in both cases material originally in the envelope is probably swept up to form the cavity.  It may be too early for the outflow to have a noticeable impact on the infall and rotation motions of the envelope.  

\noindent \textbf{Continuum emission.} The continuum emission peaks in our map at RA $=$ 18 h 30 min 04.1 s, dec. $=-02\dg\ 03\am\ 02.6\as$\ (see Extended Data Fig.~2\ref{fig:contin}) with an intensity of 93.9 mJy beam$^{-1}$, and this coincides with the centre of the \ceio\ emission (Fig.~3\ref{fig:pv_c18o}).  While the highest-intensity continuum emission (greater than $\sim50\sigma$) is concentrated and can be fit well with a two-dimensional Gaussian curve, the weaker (yet statistically significant) continuum emission is elongated northwest-southeast.  Additional continuum emission from the nearby CARMA-6 may contribute to the extended continuum emission, and molecular outflow emission is also associated with this source (although not shown here).

\clearpage
\noindent
\noindent\textbf{EXTENDED DATA}
\vspace{0.05cm}\\

\end{document}